\documentclass[prl,twocolumn,altaffilletter,nolongbibliography,numerical,flushbottom,secnumarabic,superscriptaddress,floatfix]{revtex4-2}

\usepackage{graphicx}
\usepackage{braket}
\usepackage{dsfont}  
\usepackage{comment}
\usepackage[normalem]{ulem}
\usepackage[title]{appendix}
\usepackage{amsmath,mathtools,amsthm,amssymb,pifont}
\usepackage[utf8]{inputenc}
\usepackage[american]{babel}
\usepackage{graphicx,xcolor,bbold,titlesec}
\usepackage{braket}
\usepackage{MnSymbol}

\usepackage[breaklinks, pdftex, hyperfootnotes=true, pdfpagelabels, bookmarks, pageanchor]{hyperref}
\newcommand{\bfl}{{\boldsymbol{\lambda}}}
\newcommand{\bfm}{{\boldsymbol{\mu}}}

\hypersetup{%
	colorlinks=true, linktocpage=true, pdfstartpage=1, pdfstartview=FitH, pdfborder={0 0 0},%
	breaklinks=true, pdfpagemode=UseNone, pageanchor=true, pdfpagemode=UseOutlines,%
	plainpages=false, bookmarksnumbered, bookmarksopen=true, bookmarksopenlevel=1,%
	hypertexnames=true, pdfhighlight=/O,
	urlcolor=red, linkcolor=red, citecolor=red,
	}
\usepackage{orcidlink}

\newcommand{\new}[1]{#1}

\begin{document}

\newcommand{\titleinfo}{The exact dynamical structure factor of one-dimensional hard rods and its universal random matrix behavior}
\title{\titleinfo}

\author{Oleksandr Gamayun\,\orcidlink{0000-0002-2889-8487}}
\affiliation{London Institute for Mathematical Sciences, Royal Institution, 21 Albemarle, London W1S 4BS}

\author{Mi{\l}osz Panfil\,\orcidlink{0000-0003-1525-4700}}
\affiliation{Faculty of Physics, University of Warsaw, Pasteura 5, 02-093 Warsaw, Poland}

\begin{abstract}
We obtain an exact analytic expression for the dynamical structure factor of one-dimensional quantum gas of hard rods. Our result is valid for an arbitrary many-body state of the system, with finite temperature states and the ground state being important special cases that we analyse in detail. We demonstrate that the expression obeys fundamental relations such like the f-sum rule and the detailed balance. We also reveal the hidden fermionic structure behind the correlator. In the static limit we show that it can be written  in terms of universal functions which, at zero temperature, coincide with the level spacing distribution function of the Gaussian Unitary Ensemble. Our work provides a full and exact characterisation of a dynamic correlation function in a strongly correlated interacting quantum many-body system.
\end{abstract}

\maketitle

\emph{Introduction. ---} 
The one-dimensional strongly correlated quantum many-body systems have been at the center of major developments both in equilibrium and non-equilibrium physics. Their reduced dimensionality enhances the role of correlations leading to a variety of exotic phenomena such as anomalous hydrodynamics, spin-charge separation or charge fractionalisation to name a few. Of great importance within this class are the quantum integrable models. Their exact solvability enables direct access to strongly correlated systems. A case in point is the Lieb-Liniger model~\cite{1963_Lieb_PR_130_1, 1963_Lieb_PR_130_2} which has been instrumental for both theoretical~\cite{1969_Yang_JMP_10,KorepinBOOK,2014_DeNardis_PRA_89,PhysRevA.91.023611,2017PhRvL.119s5301D,Bouchoule_2022} and experimental~\cite{kinoshita2006quantum,2009_Haller_SCIENCE_325,PhysRevX.8.021030,malvania2021generalized,kao2021topological} developments uncovering the dynamics of strongly correlated quantum matter.  

Solvability of Lieb-Liniger model gives access to its exact eigenfunctions, yet an exact determination of its dynamic correlation functions has not been achieved to date. The analytical results are limited to either large space-time regimes~\cite{1742-5468-2012-09-P09001,Shashi_2012,milosz_2021_2} or perturbative in the strength of the interactions~\cite{Cherny_2009,10.21468/SciPostPhys.9.6.082}, which motivated development of numerical methods based on the integrability~\cite{2006_Caux_PRA_74,1742-5468-2007-01-P01008,PhysRevA.89.033605,Li_2023,8qtr-dm7g}. This motivates the search for an interacting microscopic model in which such exact computations are possible. In this work, we address this problem by calculating the dynamical structure factor of a quantum gas of hard rods.

The classical gas of hard rods has been a canonical model of an interacting fluid. Its simplicity allowed for many mathematically strict developments culminated in the derivation of Navier-Stokes equations from microscopic dynamics~\cite{SpohnBook}. Among important contributions is also a determination of the density-density correlation function~\cite{flicker_pair_1968,PhysRev.171.224}.

The quantum hard rods gas has not attracted that much attention with mostly numerical studies being performed~\cite{PhysRevLett.100.020401,PhysRevA.77.043632,PhysRevA.94.043627,8gzf-v52y}. However, it has a potential to play a similar role as its classical counterpart and also complements the Lieb-Liniger gas. 
\new{In the Luttinger liquid language~\cite{GiamarchiBOOK}, the Lieb-Liniger model belongs to the phase fluctuating regime (with Luttinger liquid parameter $K>1$) while the quantum hard-rods are a microscopic model for the charge fluctuating regime ($0 < K < 1$).
Their experimental realisations were proposed in superfluid ${}^{4}{\rm He}$~\cite{PhysRevLett.95.185302,PhysRevB.94.024504} and with Rydberg atoms~\cite{PhysRevA.95.043606,Rydberg_review,8gzf-v52y}.  Hard-core interactions also naturally arise in novel experiments with polar molecules~\cite{dipolar_BEC,b8pm-3prn}.}


The quantity of interest here is the dynamic two-point correlation of the density operator,
\begin{equation} \label{S_intro}
    S(x,t) = \langle \rho_{\rm p}| \hat{\rho}(x,t) \hat{\rho}(0) |\rho_{\rm p}\rangle.
\end{equation}
evaluated in an arbitrary eigenstate of the model, here denoted by $|\rho_{\rm p}\rangle$. We compute it in a thermodynamically large system of hard rods of length $a$ and density $\rho_0$. Its Fourier transform $S(P, \omega)$ is known as the dynamic structure factor and we use both names interchangeably. Our results are valid for any values of $x$ and $t$, providing access to finite $P$ and $\omega$ regime important for scattering experiments and to large space-time asymptotics relevant for universal low-energy descriptions. The computation is based on a recently determined exact form-factor of the density operator~\cite{Kiedrzynski2025}.

\emph{Model and its spectrum. ---} The Hamiltonian of the hard-rod gas is
\begin{equation}
    H = \sum\limits_{j=1}^N \hat{p}_{j}^2 + \sum\limits_{i<j}^N V_{\rm hr}(x_i-x_j), \;\; V_{\rm hr}(x) = \begin{cases} 
        0,  &|x| > \new{a}, \\
        \infty, &|x| \leq \new{a}. 
        \end{cases}
\end{equation}
We set the mass $m=1/2$, and $\hbar=1$ which fixes the energy scale in our problem. The particles can be either fermions or bosons. The statistics of the particles influences the wavefunction but not the density-density correlation function, a situation familiar from the Tonks-Girardeau gas.

We assume system of length $L$ with periodic boundary conditions. 
The wavefunctions are Slater determinants of plane waves whose quasimomenta $\bfl = \{\lambda_j\}_{j=1}^N$ determined from the {\it Bethe} equations \cite{Nagamiya1940} are
\begin{equation}\label{Bethe}
	\lambda_j = \frac{2\pi}{L_f} \left(n_j - \nu_\bfl\right),\quad \nu_\bfl = \frac{a P_{\bfl}}{2\pi},\quad j=1,\dots N. 
\end{equation}
The quantum numbers $n_j$ should all be distinct and we introduced the reduced length $L_f \equiv L - N a$. The momentum $P_{\bfl}$ and the energy $E_{\bfl}$ of the eigenstate are
\begin{equation}  \label{energy_momentum}
	P_{\bfl} = \sum_{j=1}^N \lambda_j = \frac{2\pi}{L} \sum_{j=1}^N n_j, \qquad E_{\bfl} = \sum_{j=1}^N \lambda_j^2.
\end{equation}
A convenient way to parametrise the strength of {\it interactions} is with the help of the Luttinger parameter $K$. For the quantum hard rods $K = (1 - \rho_0 a)^2$~\cite{PhysRevLett.100.020401} where $\rho_0 \equiv N/L$ is the average density of particles~.  

The construction of the equilibrium state follows the Thermodynamic Bethe Ansatz~\cite{10.1063/1.1665585}. The particle density $\rho_{\rm p}(\lambda)$ is given by
\begin{equation} \label{rho_n}
    \rho_{\rm p}(\lambda) = \frac{1-\rho_0 a}{2\pi} n(\lambda), \quad \rho_0 = \int {\rm d}\lambda \, \rho_{\rm p}(\lambda),
\end{equation}
where the filling function $n(\lambda)$ is a standard Fermi-Dirac distribution 
\new{\begin{equation}
        n(\lambda) = \frac{1}{e^{(\lambda^2 - \mu)/T}+1},
\end{equation}
for thermal equilibrium of the temperature $T$ and the chemical potential $\mu$. 
The states that correspond to the generalized Gibbs ensembles \cite{Rigol2008} can be constructed in the same way as in other Bethe ansatz solvable models~\cite{2012PhRvL.109q5301C}.}
In the End Matter, Section~\ref{EM_Bethe}, we recall how these results follow from the quantum integrability of the hard rods.

\emph{Dynamic structure factor. ---} We consider the density-density correlation~\eqref{S_intro} in a finite system of $N$ particles, of length $L$ and in state $| \bfl \rangle$. We shall think about this state as a representative state of a thermodynamic state $|\rho_p\rangle$ with density $\rho_{\rm p}(\lambda)$ such that 
\begin{equation}\label{ff0}
    S(x,t) = \lim_{\rm th} \,\langle \bfl|\hat{\rho}(x,t)\hat{\rho}(0) | \bfl \rangle,
\end{equation}
with the usual notion of the thermodynamic limit $N/L~\rightarrow~\infty $ with $\rho_0 \equiv N/L$ fixed. 
The  form factor (spectral) expansion of it reads 
\begin{equation}\label{ff1}
       S(x,t) =  \lim_{\rm th} \sum_{\bfm} e^{it (E_{\bfm}-E_{\bfl})-ixP_{\bfm}} |\langle \bfm | \rho| \bfl \rangle|^2.
\end{equation}
The summation here is performed over all possible sets of $N$ distinct quantum numbers $\{n_j\}$ that determine the quasimomenta $\{\mu_j\}$ according to the Bethe equations~\eqref{Bethe}. The energy and momentum are given by~\eqref{energy_momentum}.
The density form factors have been recently determined~\cite{Kiedrzynski2025}. Up to a phase factor irrelevant to the dynamic structure factor (DSF), they are
\begin{equation}\label{ff22}
    \langle \bfm | \rho| \bfl \rangle = \frac{\Delta P}{L} \left(\frac{2\sin \left(a \Delta P/2\right)}{L_f} \right)^{N-1} C[\bfm, \bfl],
\end{equation}
where $\Delta P = P_\bfm - P_\bfl$ and $C[\bfm, \bfl]$ is the Cauchy determinant
\begin{equation}
    C[\bfm, \bfl] = \det\left(\frac{1}{\mu_i - \lambda_j} \right). 
\end{equation}

The evaluation of the DSF involves two technical difficulties. First, it requires a sum over (squares of) Cauchy determinants. Second, the sums over different quasimomenta are not independent. As seen from the Bethe equations~\eqref{Bethe}: modifying a single quantum number changes not only the corresponding quasimomentum but also all the other quasimomenta through the collective dressing encoded by factors $\nu_\bfm$.

\begin{figure}[t]
    \centering
    \includegraphics[width=\linewidth]{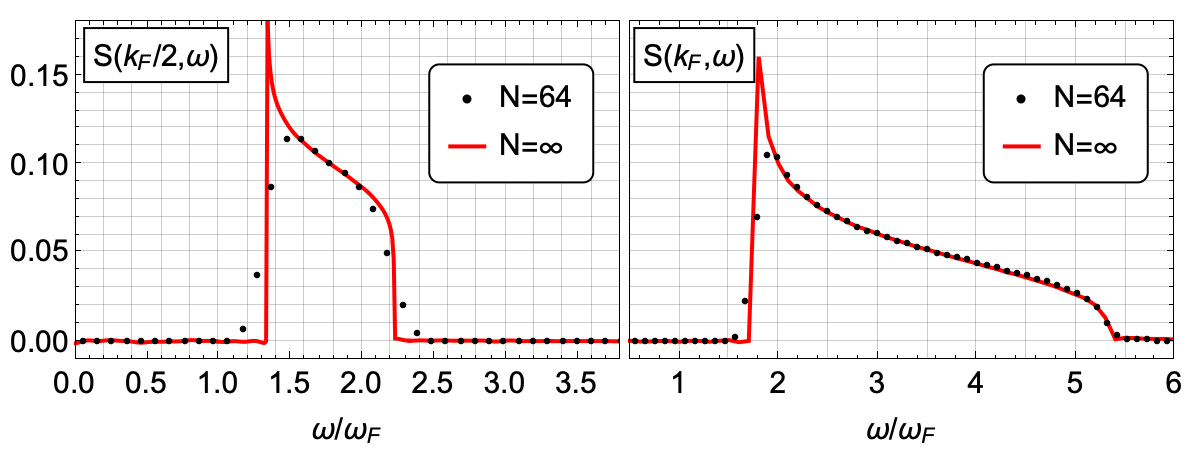}
    \caption{The comparison of the Fourier transformed DSF in the thermodynamic limit and the finite system for zero temperature, $k_F=\pi \rho_0$, and $a\rho_0=1/4$.}
    \label{Fig000}
\end{figure}

The first problem would be solvable if factors $\nu_\bfm$ were in fact constant. The mathematical task becomes then very similar to that of the integrable mobile impurity 
\cite{Gamayun2015,Gamayun2016,impurity_us,10.21468/SciPostPhys.17.2.063} and the corresponding methods can be adapted to the present situation. This allows us to compute the spectral sum in the thermodynamic limit for a fixed $\nu$. To account for the constraint $\nu = \nu_{\bfm}$ we consider all values of $\nu$ and subsequently filter out the contributions from $\nu \neq \nu_\bfm$. The details are presented in Sec.~\ref{EM_Fredholm} of the End Matter.

The result for the dynamic density-density correlation function is 
\begin{equation}\label{Sxt}
    S(x,t) =\int\limits_{-\infty}^{\infty} \frac{dP}{2\pi} \, \frac{ \sqrt{K} P^2e^{-i P x}  }{\left(2\sin (\pi \nu)\right)^2 } \, 
    \int\limits_{-\infty}^{\infty}\, ds \,e^{is    P} \mathcal{D}_\nu(s,t),
\end{equation}
here $\nu = aP/(2\pi)$ and $\mathcal{D}_\nu(x,t)= \det(1+n \hat{V})$ is a Fredholm determinant of the operator $\hat{V}$ that acts on the real line. It is weighted with the filling function $n(\lambda)$ and has the following explicit kernel 
\begin{equation}\label{kernel}
    V(\lambda,\mu) = \frac{1}{\new{\pi}} \frac{e_+(\mu)e_-(\lambda)-e_+(\lambda)e_-(\mu)}{\mu- \lambda}
\end{equation}
with $e_-(\lambda) = e^{-it\lambda^2/2+i\lambda s/2}$ and 
\begin{multline}
    e_+(\lambda)  = - \sin(\pi \nu) e^{it \lambda^2/2 - i \lambda s/2} \\\times \left[
\cos(\pi \nu) + i  \sin(\pi \nu)  {\rm Erf} \left(\frac{s-2\lambda t}{2\sqrt{-it}}\right)
\right]
\end{multline}
The numerical evaluation of the Fredholm determinant can be easily implemented by the quadrature method~\cite{Bornemann2009}.

\emph{Basic properties. ---} The exact form of the kernel allows us to conclude the following involution properties of the determinant
\begin{equation}\label{invo}
    \mathcal{D}_{\nu}(s,t) = \mathcal{D}_{-\nu}(-s,t) =  \mathcal{D}^*_{-\nu}(s,-t) =  \mathcal{D}^*_{\nu}(-s,-t).
\end{equation}
Due to these properties 
the DSF in energy-momentum space, 
\begin{equation}\label{Spw}
    S(P,\omega)=  \int\limits_{-\infty}^{\infty}ds \int\limits_{-\infty}^{\infty} dt\,\frac{ \sqrt{K} P^2 e^{isP -i\omega t} }{\left(2\sin (\pi \nu)\right)^2 } \mathcal{D}_\nu(s,t), 
\end{equation}
is real and symmetric w.r.t. to the momentum flip, i.e. $S(P,\omega) = S(-P,\omega)$. 

In Fig.~\ref{Fig000} we show the fixed momentum cuts for different values of the momentum. We choose the representative value of $a \rho_0 = 1/4$ and compare with the finite system results obtained in the previous work~\cite{Kiedrzynski2025}. The finite size results visually converge towards the thermodynamic limit everywhere except the edge singularity where significantly larger system sizes are required to obtain thermodynamic behavior. 



Moreover, using methods of Ref.~\cite{PhysRevResearch.5.043265} we can also demonstrate that for the thermal distribution with the temperature $T$ there is an additional analytic property 
\begin{equation}
    \mathcal{D}_\nu(s,t) = \mathcal{D}_\nu(s, -t + i/T). 
\end{equation}
This implies the detailed balance condition 
\begin{equation}
    S(P,-\omega) = S(P,\omega)e^{-\omega/T },
\end{equation}
and in the real spaces the \new{Kubo-Martin-Schwinger} relation
\begin{equation}
    S(x,t) = S(x, -t + i/T).
\end{equation}
Another key property is the $f$-sum rule fixing the frequency integrated DSF
\begin{equation}\label{N}
        \int \frac{d\omega}{2\pi} \omega S(P,\omega) =   \rho_0 P^2. 
\end{equation}
The integral over the frequencies in the right hand side of~\eqref{Spw}, 
results in the time derivative of the Fredholm determinant at $t=0$. This expression can be computed explicitly and turns out to be proportional to the delta function $\delta(s)$, afterwards the result \eqref{N}, follows.

Finally, we consider the static covariance of the density operator
\begin{equation} \label{cov_mat}
    C = \int {\rm d}x \, \left( \langle \hat{\rho}(x,0) \hat{\rho}(0) \rangle - \langle \hat{\rho}(0) \rangle^2 \right).
\end{equation}
This quantity is a special case of the static covariance matrix defined for any pair of local conserved densities. It plays an important role in transport~\cite{milosz_2021_2} and can be computed directly from the thermodynamics of the quantum integrable model~\cite{Doyon_drude}; for the quantum hard rods gas it reads
\begin{equation} \label{cov_qhr}
    C = (1 - \rho_0 a)^3 \int \frac{{\rm d}k}{2\pi} n(k)(1-n(k)).
\end{equation}
We have verified that $S(x,t)$ given by Eq.~\eqref{Sxt} satisfies this relation. The details of this and other computations of this section are presented in the End Matter Sec.~\ref{EM_details}. 

These general identities validate the expression for the dynamic structure factor and show that it is not merely formal, but admits further analytical manipulation from which physical insights can be obtained. In the following, we reveal how the DSF of quantum hard rods relates to a free fermionic and random matrix theory structures.

 \begin{figure}
    \includegraphics[scale=0.4]{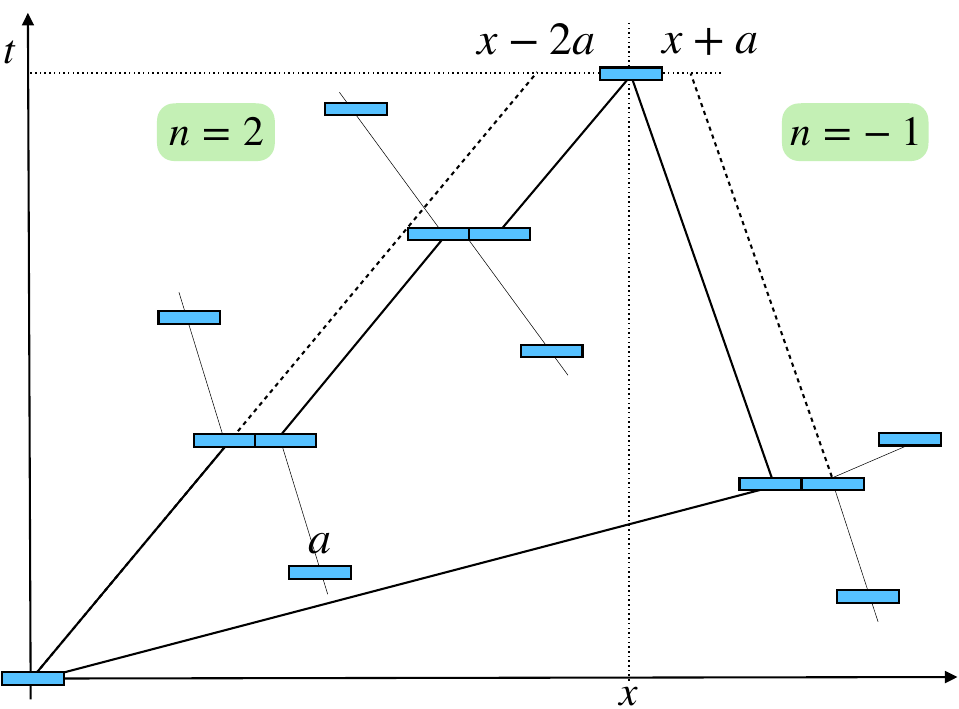}
    \caption{Function $S_n(x,t)$ for $n>0$ can be interpreted as contribution to the correlation function involving scattering with $|n|$ particles. Here two sources of the correlation between $(0,0)$ and $(x,t)$ are shown. 1) for $n=2$, a particle starting at $(0,0)$ scatters with two other particles. The signal reaches at time $t$ point $x$ for hard rods of length $a$ and point $x-2a$ for hard-core particles (dashed line). 2) for $n=-1$ the scattering is at position larger than $x$ and the signal reaches point $x$ or $x+a$ for hard rods and hard-core particles respectively. This explains the shift in the argument $S_n(x - na, t)$ in~\eqref{Sxt11}.}.
    \label{fig:hr_corr}
\end{figure}

\emph{Free fermionic structures. ---}  An equivalent presentation for the DSF can be obtained by employing the fact that the determinant is periodic in $\nu$, $\mathcal{D}_{\nu+1}=\mathcal{D}_{\nu}$. Let us integrate twice by parts in $s$ in \eqref{Sxt} to remove $P^2$. Afterwards, we restrict integration in $P$ to the fundamental domain $ [0,2\pi/a]$ and take into account that
\begin{equation}
    \sum\limits_{n\in \mathds{Z}} e^{i2\pi(s-x)n/a} = a     \sum\limits_{n\in \mathds{Z}}  \delta(s-x+na),
\end{equation}
to arrive at the following presentation for the DSF
\begin{equation}\label{Sxt11}
    S(x,t)= \sqrt{K} \sum\limits_{n\in \mathds{Z}} S_n(x-na,t),
\end{equation}
where 
\begin{equation}\label{sn}
      S_n(x,t) =  -  \int\limits_{0}^{1}  \frac{e^{-2\pi i \nu n }  }{4\sin^2(\pi \nu)} \partial_x^2\mathcal{D}_{\nu}(x,t)d\nu .
\end{equation}
This way, we have traded the integral over $s$ in \eqref{Sxt} for the infinite sum. Functions $S_n(x,t)$ are independent of the rods length $a$. They can be interpreted as contributions from scattering with $|n|$ particles in a theory of hard-core particles (free fermions or Tonks-Girardeau gas), see Fig.~\ref{fig:hr_corr}. Crucially, all the complexity of the dynamic correlation function of the hard rods as compared with hard-core particles, arises because of different mixing of the contributions $S_n(x,t)$ in~\eqref{Sxt11}.

It is also interesting to notice that the contribution for $n=0$ is nothing but a 
continuous version of the spin-spin correlation function for the Hubbard model in strong coupling limit (see Eq. (56) in \cite{10.21468/SciPostPhys.15.2.073}).

\emph{The static limit and the random matrix theory. ---}  The presentation \eqref{Sxt11} turns out to be especially fruitful for the static limit $S(x) = S(x,t=0)$. In this case the kernel \eqref{kernel} transforms into the sine-kernel 
\begin{equation}
    \hat{V}\Big|_{t=0} =(e^{2\pi i \nu }-1)\hat{\mathcal{K}},\quad \mathcal{K}(q,p) =  \frac{\sin \frac{s(q-p)}{2}}{\pi(q-p)}.
\end{equation}
This expression is valid for positive $s$, while for $s<0$ the kernel is obtained by complex conjugation (see \eqref{invo}). Consequently, the second derivative in \eqref{sn} must be handled with particular care. In particular, the discontinuity of the second derivative gives rise to terms proportional to \new{$\rho_0\delta(x)$}. In the following, we restrict our attention to the region $x>0$, where no such issues arise.

\begin{figure}
    \centering
    \includegraphics[width=0.9\linewidth]{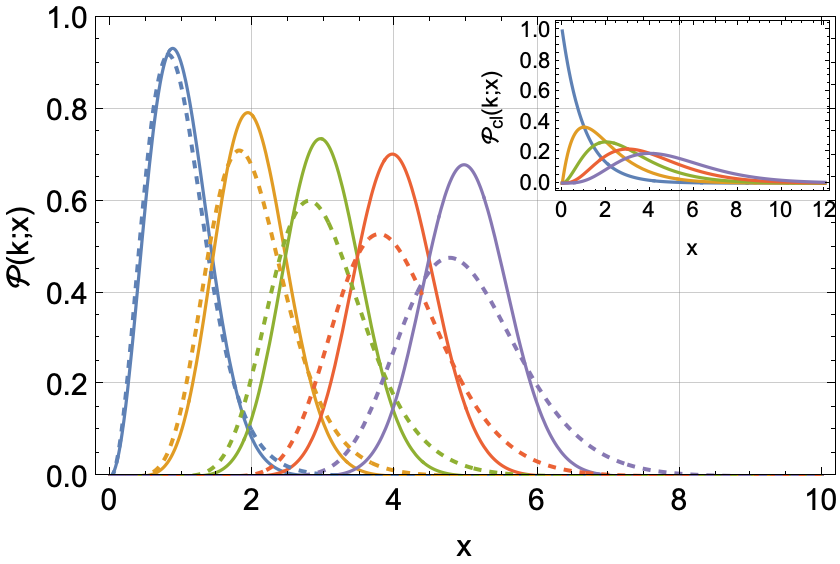}
    \caption{Plots of the k-level spacing probability densities ($k=0,1,2,3,4$) for the zero temperature (solid lines), and for $T=10$, $a=0.5$, $\rho_0=1$  (dashed lines); The inset shows the classical result \eqref{pcl}.}
    \label{FigDISTR}
\end{figure}

We notice that for $n \in \mathds{Z}$, 
\begin{equation}
     \int\limits_{0}^{1} \frac{d\nu}{\sin^2(\pi \nu)} (e^{2\pi i \nu}-1)^m e^{2i\nu \pi |n|}  = 0 ,\qquad m\ge 2. 
\end{equation} 
Therefore, in the sum \eqref{Sxt11} the only terms that survive have the same signs of $n$ and $x-na$. This means that for $x>0$ the summation will be from $n=1$ to $n = [x/a]$ - the greatest number of the rods one can fully allocate in $[0,x]$. 
The computation of integrals in \eqref{sn} reduces to evaluating residues, which are expressed via the derivatives of the Fredholm determinant. To present answers compactly, we introduce 
\begin{equation}\label{E}
   E(m;x) \equiv \frac{\partial_{\xi}^{m}}{m!}\det\left(1 + (\xi-1)n \hat{\mathcal{K}}\right)\Big|_{\xi =0 } 
\end{equation}
and 
\begin{equation}\label{p}
    \tilde{p}(k;x) =\partial_x^2 \sum\limits_{j=0}^{k}(k+1-j)E(j;x). 
\end{equation}
At zero temperature these functions are closely related to the correlation functions in the Gaussian Unitary Ensemble \cite{forrester_log-gases_2010}.  Indeed, $E(m,x)$ plays the role of the probability of $m$ eigenvalues in the interval of the length proportional to $x$, and $p(k,x)$ gives the probability density of the $k$ successive eigenvalues in the interval $[x, x + {\rm d}x]$. To be more precise, we have to take into account the appropriate normalization, which for $\tilde{p}$ reads 
\begin{equation}
    \int\limits_{0}^\infty \tilde{p}(k;x)  dx =\mathcal{N}, \qquad \mathcal{N} = \frac{\rho_0}{1 - \rho_0 a}.
\end{equation}
This way, we introduce $\mathcal{P}(k;x) = \mathcal{N}^{-2}\tilde{p}(k;x/\mathcal{N})$, which have proper normalization and the mean value
\begin{equation}
    \int\limits_{0}^\infty \mathcal{P}(k;x)  dx = 1,\qquad 
     \int\limits_{0}^\infty x \mathcal{P}(k;x)  dx = k+1.
\end{equation}
The functions $\mathcal{P}(k;x)$ can be easily tabulated numerically for any reasonable distribution $n(\lambda)$ (see for example \cite{Bornemann2010numerical}). The first five of these functions are presented in Fig.~\ref{FigDISTR}. 
Finally, with all the functions now clearly defined, we can carry out the remaining algebra and obtain the expression for the static structure factor
\begin{equation}\label{sx}
    S(x) = \frac{\rho_0^2}{1-\rho_0 a} \sum\limits_{n=1}^{[x/a]} \mathcal{P}\left(n-1;\frac{\rho_0(x-n a)}{1-\rho_0 a}\right).
\end{equation}
It is interesting to notice that the classical answer has the same form but with \new{$\mathcal{P}$ being a Poisson distribution}~\cite{flicker_pair_1968}
\begin{equation}\label{pcl}
    \mathcal{P}(k;x) \to \mathcal{P}_{\rm cl}(k;x) =  \frac{x^k}{k!} e^{-x}.
\end{equation}
The classical correlations of the observables that do not involve momenta do not depend on temperature. 
This way, to formally recover \eqref{pcl} from the quantum answer we send $T\to\infty$. In this limit the determinant in \eqref{E} simplifies as  $\det(1+(\xi-1) n\mathcal{K})) \approx e^{(\xi-1){\rm Tr}n\mathcal{K}}$, which gives 
$E(m;x) \approx (x\mathcal{N})^m e^{-x\mathcal{N}}/m!$. Substituting this result into \eqref{p} and performing summation along with the rescaling leads to \eqref{pcl}.
\begin{figure}
    \centering
    \includegraphics[width=0.9\linewidth]{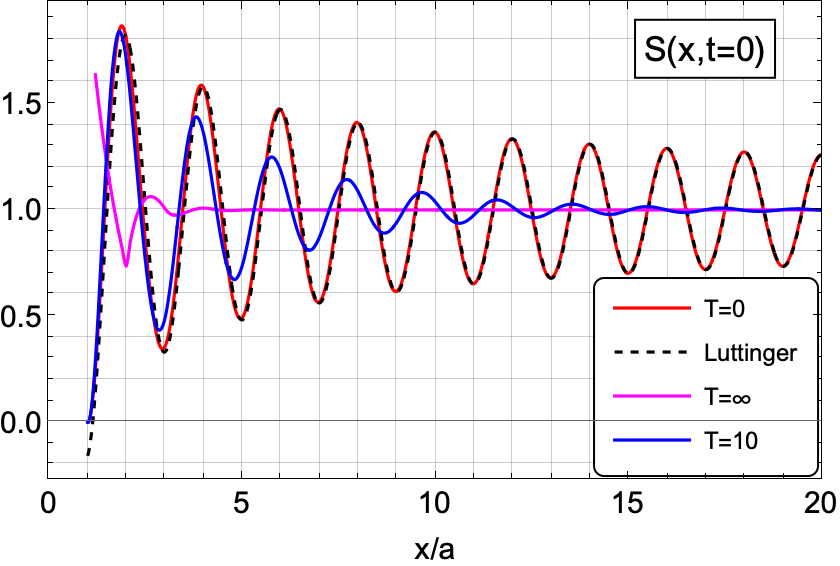}
    \caption{The static structure factor at $\rho_0=1$, $a=0.5$, and the various temperature shown in the legends. The classical answer (magenta line) corresponds to \eqref{pcl}. The "Luttinger" is the answer obtained in the Luttinger model approximation in Eq. \eqref{Luttinger}.}
    \label{Fig2}
\end{figure}

The typical behavior of $S(x)$ is shown in Fig.~\ref{Fig2}. We see the pronounced Friedel osciallations, whose amplitude strongly depends on the rod size $a$. As $a\to 0$ the amplitude vanishes, and the whole expression takes the following form
\begin{equation}\label{ato0}
    S(x) \overset{a\to 0}{=} \rho_0^2 - \left|\frac{1}{2\pi}\int e^{-ipx} n(p)dp\right|^2.
\end{equation}
The details of the classical and free fermionic limit are presented in Section~\ref{EM_details} of the End Matter.  

In the opposite limit $\rho_0 a \to 1$, Eq. \eqref{sx} suggests that the structure factor turns into the sum of delta functions, which reflects the {\it crystallization} of the system due to the dense pack of the hard-rods. 
The dashed line in Fig. \eqref{Fig2} is the prediction from the Luttinger model~\cite{GiamarchiBOOK},
\begin{equation}\label{Luttinger}
\frac{S(x)}{\rho_0^2} = 1 - \frac{K}{2(k_F x)^2} + \sum\limits_{m=1}^\infty \frac{A_m\cos(2mk_F x)}{(\rho_0 x)^{2m^2K}},
\end{equation}
\new{with the nonuniversal prefactors $A_m$ determined in~\cite{Kiedrzynski2025}.}
As we see the predictions are pretty good, however, the smaller the $K$ the more terms once should account for in the sum \eqref{Luttinger}.
The temperature leads to the faster decay of the oscillations which is expected from the Luttinger theory.

\emph{Conclusions. ---}
We have obtained an exact expression for the dynamic structure factor of a gas of quantum hard rods. The result is valid for {\em any} state of the system characterised by the quasimomenta distribution $\rho_{\rm p}(\lambda)$ making it applicable to various physical situations. We have considered  in detail the case of thermal equilibrium. At zero temperature we have found interesting relations with the random matrix theory, while at large temperatures we have recovered correlations of the classical hard rods gas.

Furthermore, the integrability of the model yields a non-perturbative control over the interaction parameter. This provides information on dynamic correlations in both weakly correlated case, $a \rho_0 \approx 0$, resembling the free fermionic gas up to to the tightly packed, strongly correlated system for $a \rho_0 \approx 1$ exhibiting {\em crystallization}. 

Finally, the formula is valid also for arbitrary value of $x$ and $t$ (or, equivalently any value of $P$ and $\omega$) providing access to features of the correlator that are not captured by universal low-energy descriptions such as Luttinger liquid theory or hydrodynamics more generally. This includes the regime probed in scattering experiments. 

Altogether, this establishes the results of this work as an important benchmark for theoretical and experimental studies of strongly correlated quantum systems and opens new research directions, for example, towards deriving the linear and non-linear Luttinger liquid theory from the exact microscopic description.  

\begin{acknowledgments}
\paragraph{Acknowledgments. ---}  
We are grateful to Vir Bulchandani, Krzysztof Jachymski, Gr{\'e}gory Schehr and Herbert Spohn for inspiring discussions. 
O.G. acknowledges support from Cognia AI. 
M.P. acknowledges support from the National Science Centre, Poland, under the OPUS grant 2022/47/B/ST2/03334.
\end{acknowledgments}

\vspace{0.25cm}

\noindent {\bf Data Availability:} All the data required to reproduce the plots are available at~\cite{dataQHR}.

\bibliographystyle{apsrev4-2}
\bibliography{literatura}
\clearpage
\setcounter{section}{0}
\setcounter{secnumdepth}{2}

\begin{center}
    \textbf{End Matter}
\end{center}

\section{Coordinate and Thermodynamic Bethe Ansatz} \label{EM_Bethe}
The scattering phase shift of hard rods potential $V_{\rm HR}(x)$ is $\theta(\lambda) = -a\lambda $. The Bethe equations 
\begin{equation}\label{Bethe_generic}
    \lambda_j = \frac{2\pi}{L} n_j +\frac{1}{L} \sum\limits_{l\neq j}^N \theta(\lambda_j - \lambda_l)
\end{equation}
can be in this case solved, with the solution presented in~\eqref{Bethe}. In a similar fashion, the Thermodynamic Bethe Ansatz simplifies. The non-linear equation for the pseudoenergy (dispersion) $\epsilon(\lambda)$ is 
\begin{equation}
    \epsilon(\lambda) = \epsilon_0(\lambda) + \frac{1}{2\pi}\int {\rm d}\mu K(\lambda - \mu) \log\left( 1 + \exp(-\epsilon(\mu))\right),
\end{equation}
with $K(\lambda) = \partial \theta(\lambda) /\partial \lambda$. For hard rods $K = -a$ and the solution takes form $\epsilon(\lambda) = \epsilon_0(\lambda) + {\rm const}$. Thus the only effect of the interactions is in the shift of the chemical potential. 

Finally, the particles density $\rho_{\rm p}(\lambda) = \rho_{\rm tot}(\lambda) n(\lambda)$ where $n(\lambda) = 1/(1 + e^{\epsilon(\lambda)})$ and $2\pi \rho_{\rm tot}(\lambda) = (1)^{\rm dr}(\lambda)$. The dressing of a function $f$ is defined as solution to the following integral equation
\begin{equation} \label{dressing}
    f^{\rm dr}(\lambda) = f(\lambda) + \int {\rm d}\mu K(\lambda - \mu) n(\mu) f(\mu).
\end{equation}
For the hard rods this gives $2\pi\rho_{\rm tot}(\lambda) = 1 - \rho_0 a$ which gives particle distribution $\rho_{\rm p}(\lambda)$ reported in Eq.~\eqref{rho_n}.

\section{Fredholm determinant presentation} \label{EM_Fredholm}

The spectral sum in Eq. \eqref{ff1} is taken over the set of $N$, ordered integers $n_1<n_2\dots < n_N$ that provide unique ordered sets of $\bfm = \{\mu_1,\mu_2,\dots \mu_N\}$ with
\begin{equation}
	\mu_j = \frac{2\pi}{L_f} \left(n_j - \nu_\bfm\right),\qquad \nu_\bfm = \frac{a P_{\bfm}}{2\pi}.
\end{equation}
To disentangle the quasimomenta we employ the following trick: first, we identically present  
\begin{equation}
    \sum\limits_{\bfm} (\dots) =     \sum\limits_{\bfm} \sum_P \int\limits_{-L_f/2}^{L_f/2} \frac{ds}{L_f} \, e^{is (   P- \sum\limits_{j=1}^N \mu_j )} (\dots)
\end{equation}
Here, the integral over $s$ singles out the admissible sets. Now we can interchange the summation order and treat each $\mu_j \in \bfm$ as an independent variable (restricted only by the Pauli principle) with $\nu = a P /(2\pi)$. 
This way, the spectral sum of~\eqref{ff1} becomes 
\begin{equation}
    S(x,t) =\int\limits_{-L_f/2}^{L_f/2} \frac{ds}{L_f}\sum_{P,\bfm} e^{it (E_{\bfm}-E_{\bfl})-i(x+s)P_{\bfm}+iPs}  |\langle \bfm | \rho| \bfl \rangle|^2.
\end{equation}

The summation over $\bfm$ can be taken, and in the thermodynamic limit is expressed in terms of the Fredholm determinant 
\begin{equation}
    \sum\limits_{\bfm} e^{it (E_{\bfm}-E_{\bfl})-is P_{\bfm}} C^2[\bfl,\bfm]= \det(1+n\hat{V}).
\end{equation}
with the kernel defined in \eqref{kernel}  ({\it c.f.} \cite{Gamayun2016}). For convenience we have additionally shifted $s\to s-x$. Final expression~\eqref{Sxt} for $S(x,t)$ follows from taking the thermodynamic limit completely replacing
\begin{equation}
    \sum_P = \frac{L}{2\pi} \int dP.
\end{equation}

\section{Details of other derivations} \label{EM_details}

{\bf f-sum rule:}
To evaluate the $f$-sum rule \eqref{N} we use presentation \eqref{Spw}, which gives 
\begin{equation}
    \int \frac{d\omega}{2\pi} \omega S(P,\omega) =
     \frac{(-i) \sqrt{K} P^2  }{\left(2\sin (\pi \nu)\right)^2 }   \int\, ds \,e^{is    P}   \partial_t\mathcal{D}(s,t)\Big|_{t=0}.
\end{equation}
The derivative of the kernel at $t=0$ reads 
\begin{align}\nonumber
     \partial_t V(\lambda,\mu)\Big|_{t=0} = \frac{i(\mu+\lambda)}{2\pi} &\left(e_+(\mu)e_-(\lambda)+e_+(\lambda)e_-(\mu)\right) \\&+ \frac{2\sin^2(\pi \nu)}{\pi} i\delta(s)
\end{align}
Since all the terms in this expression are kernels of  rank-1, the derivative of the determinant reads 
\begin{multline}
    \label{dt}
    \partial_t \mathcal{D}(s,t)\Big|_{t=0} =\det(1+n\hat{V}+n\hat{V}_{+}) -    \mathcal{D}(s,t) + \\ \det(1+n\hat{V}+n\hat{V}_{-}) -    \mathcal{D}(s,t) + 
    4i\sin^2(\pi \nu) \mathcal{N}\delta(s)
\end{multline}
with 
\begin{equation}
    V_{\pm}(\lambda,\mu) = \frac{e^{2\pi i \nu}-1}{\pi} p e^{\pm is(\mu-\lambda)/2 }.
\end{equation}
Notice that all the determinants are defined on the symmetric interval so under the change $\lambda\leftrightarrow -\lambda$, $\mu\leftrightarrow -\mu$ we obtain that $V_+ \to - V_-$, so the whole combination \eqref{dt} vanishes apart for the last term. This term after the integration over $s$ gives \eqref{N}. 

{\bf The static covariance matrix:}
\label{sumM}
To compute the static covariance matrix~\eqref{cov_mat} we rewrite it as 
\begin{equation}
    C = -2 \int\limits_0^\infty x \partial_x S(x) dx 
\end{equation}
Then we use presentation \eqref{Sxt} for $S(x) \equiv S(x,t=0)$. The integration over $x$ produces the derivative of the $\delta$-function in $P$, which renders the whole expression into 
\begin{equation}
   C =  2\sqrt{K} {\rm Re} \lim\limits_{P\to 0}   \partial_P\int\limits_{0}^{\infty}\frac{P^3 e^{is    P}  \,\mathcal{D}(s,\nu) }{\left(2\sin (\pi \nu)\right)^2 }\, ds
\end{equation}
The integral is mostly accumulated on the large distances, so we can use the asymptotic expansion of the determinant as $s\to\infty$ \cite{DannyCherno}
\begin{equation}
    \mathcal{D}(s,\nu) \approx \exp\left(s\int \frac{d\lambda}{2\pi}\ln\left(1 + n(\lambda) (e^{2\pi i\nu}-1)\right) \right)
\end{equation}
This way, integral over $s$ can be easily computed 
\begin{equation}
     \int\limits_{0}^{\infty}e^{is    P}  \,\mathcal{D}(s,\nu) \, ds  \approx \frac{i P^{-1}}{1+ a \Xi_0 } + \frac{ a^2 \Xi_2}{2 (1+ a \Xi_0 )^2} + O(P) 
\end{equation}
where we introduced
\begin{equation}
    \Xi_0 = \int\limits_{-\infty}^{\infty} \frac{n(\lambda){\rm d}\lambda}{2\pi} ,\quad
    \Xi_2 =  \int\limits_{-\infty}^{\infty} \frac{n(\lambda)(1-n(\lambda)){\rm d}\lambda}{2\pi} .
\end{equation}
This way, we obtain 
\begin{equation}
C=  \sqrt{K}\frac{\Xi_2}{(1+ a \Xi_0)^2}.
\end{equation}
From the TBA it follows that $\Xi_0 = \rho_0/(1- a\rho_0)$ and  we recover Eq.~\eqref{cov_qhr}.

On the other hand, formula~\eqref{cov_qhr} follows from a specialization to the case of quantum hards of the general expression~\cite{Doyon_drude}
\begin{equation}
    C = \int {\rm d}\lambda\, \rho_p(\lambda)(1 - n(\lambda)) h_0^{\rm dr}(\lambda) h_0^{\rm dr}(\lambda),
\end{equation}
where $h_0(\lambda) = 1$ for galilean invariant systems. The dressing operation~\eqref{dressing} for the quantum hard rods results in multiplying by factor $(1 - a \rho_0)$. Using then the relation between particle distribution $\rho_{\rm p}(\lambda)$ and the filling function $n(\lambda)$
leads to~\eqref{cov_qhr} of the main text.

{\bf Classical limit:}
The classical limit formally corresponds to the infinitely large temperatures. In this case we can approximate the determinant in \eqref{E} as 
\begin{equation}
\det(1+(\xi-1) n\mathcal{K}) \approx e^{(\xi-1){\rm Tr}n\mathcal{K}}  = e^{(\xi-1)x\mathcal{N}}
\end{equation}
This way, $E(m;j)\approx y^m e^{-y}/m!$, with $y=x \mathcal{N}$. This leads to the following expression for the probability density in~\eqref{p}
\begin{equation}
\frac{\tilde{p}(k;x)}{ \mathcal{N}^2 } =e^{-y}\sum\limits_{j=0}^k (k+1-j) \left(\frac{y^{j-2}}{(j-2)!} -\frac{2y^{j-1}}{(j-1)!} +\frac{y^{j}}{j!}  \right).
\end{equation}
Shifting the summation index, we obtain 
\begin{multline}
    \frac{\tilde{p}(k;x)}{ \mathcal{N}^2 } =\sum\limits_{j=0}^{k-2} (k-1-j) \frac{y^{j}e^{-y}}{j!}   -2 \sum\limits_{j=0}^{k-1} (k-j) \frac{y^{j}e^{-y}}{j!}\\
+\sum\limits_{j=0}^{k} (k+1-j) \frac{y^{j}e^{-y}}{j!}  = \frac{y^{k}e^{-y}}{k!} 
\end{multline}
After the rescaling $ \mathcal{P}(k;x)=\mathcal{N}^{-2}\tilde{p}(k;x/\mathcal{N})$ we recover~\eqref{pcl}.

{\bf Free-fermionic limit:}
Let us analyze the free-fermionic case by taking $a\to 0$ for the static structure factor. 
Using Eqs. \eqref{Sxt11} and \eqref{sn} for $t=0$, we 
present 
\begin{equation}
    S(x) \overset{a\to 0}{=} - \int\limits_{0}^{1}  \frac{\sum\limits_{n=1}^\infty e^{-2\pi i \nu n }  }{4\sin^2(\pi \nu)} \partial_x^2\mathcal{D}_{\nu}(x,t)d\nu
\end{equation}
Let us denote $\xi = e^{2\pi i \nu}$, for convergence, we have to assume that its absolute value is slightly bigger than one. In this case, performing summation we obtain 
\begin{equation}
    S(x) \overset{a\to 0}{=}  \oint\limits_{|\xi|=1+0} \frac{d\xi}{2\pi i}  \frac{\partial_x^2 \det(1+(\xi-1)n\hat{\mathcal{K}})}{(\xi-1)^3}
\end{equation}
Taking into account that 
\begin{multline}
    \det(1+(\xi-1)n\hat{\mathcal{K}}) = 1 + (\xi-1)\int n(p)\mathcal{K}(p,p) dp \\ + \frac{(\xi-1)^2}{2}\int \left|
    \begin{array}{cc}
        n(p)\mathcal{K}(p,p) & n(p)\mathcal{K}(p,q) \\
        n(q)\mathcal{K}(q,p) & n(q)\mathcal{K}(q,q)
    \end{array}
    \right|dq dp \\+ O((\xi-1)^3).
\end{multline}
We can easily compute the residue at $\xi=1$. Afterwards, we arrive at Eq. \eqref{ato0}.

\end{document}